\documentstyle[12pt]{article}
\newcommand\mpl{M_{\rm P}}

\catcode`\@=11
\def\marginnote#1{}
\def\ifmath#1{\relax\ifmmode #1\else $#1$\fi}

\def\identity{1 \hspace{-.085cm}{\rm l}}

\def\bold#1{\setbox0=\hbox{$#1$}%
     \kern-.025em\copy0\kern-\wd0
     \kern.05em\copy0\kern-\wd0
     \kern-.025em\raise.0433em\box0 }

\def\GENITEM#1;#2{\par\vskip6pt \hangafter=0 \hangindent=#1
   \Textindent{$ #2$ }\ignorespaces}

\newcount\hour
\newcount\minute
\newtoks\amorpm
\hour=\time\divide\hour by60
\minute=\time{\multiply\hour by60 \global\advance\minute by-
\hour}
\edef\standardtime{{\ifnum\hour<12 \global\amorpm={am}%
    \else\global\amorpm={pm}\advance\hour by-12 \fi
    \ifnum\hour=0 \hour=12 \fi
    \number\hour:\ifnum\minute<100\fi\number\minute\the\amorpm}}
\edef\militarytime{\number\hour:\ifnum\minute<100\fi\number\minute}
\def\draftlabel#1{{\@bsphack\if@filesw {\let\thepage\relax
  \xdef\@gtempa{\write\@auxout{\string
    \newlabel{#1}{{\@currentlabel}{\thepage}}}}}\@gtempa
    \if@nobreak \ifvmode\nobreak\fi\fi\fi\@esphack}
     \gdef\@eqnlabel{#1}}
\def\@eqnlabel{}
\def\@vacuum{}
\def\draftmarginnote#1{\marginpar{\raggedright\scriptsize\tt#1}}
\def\draft{\oddsidemargin -.5truein
        \def\@oddfoot{\sl preliminary draft \hfil
        \rm\thepage\hfil\sl\today\quad\militarytime}
        \let\@evenfoot\@oddfoot \overfullrule 3pt
        \let\label=\draftlabel
        \let\marginnote=\draftmarginnote

\def\@eqnnum{(\theequation)\rlap{\kern\marginparsep\tt\@eqnlabel}%
\global\let\@eqnlabel\@vacuum}  }
\def\preprint{\twocolumn\sloppy\flushbottom\parindent 1em
        \leftmargini 2em\leftmarginv .5em\leftmarginvi .5em
        \oddsidemargin -.5in    \evensidemargin -.5in
        \columnsep 15mm \footheight 0pt
        \textwidth 250mmin      \topmargin  -.4in
        \headheight 12pt \topskip .4in
        \textheight 175mm
        \footskip 0pt

\def\@oddhead{\thepage\hfil\addtocounter{page}{1}\thepage}
        \let\@evenhead\@oddhead \def\@oddfoot{} \def\@evenfoot{}
}
\def\titlepage{\@restonecolfalse\if@twocolumn\@restonecoltrue\o
necolumn
     \else \newpage \fi \thispagestyle{empty}\c@page\z@
        \def\thefootnote{\fnsymbol{footnote}} }
\def\endtitlepage{\if@restonecol\twocolumn \else  \fi
        \def\thefootnote{\arabic{footnote}}
        \setcounter{footnote}{0}}  
\catcode`@=12
\relax
\def\be{\begin{equation}}
\def\ee{\end{equation}}
\def\bea{\begin{eqnarray}}
\def\eea{\end{eqnarray}}
\def\simlt{\stackrel{<}{{}_\sim}}
\def\simgt{\stackrel{>}{{}_\sim}}

\def\mst11{m_{\;\\widetilde{t}_{1}}}

\def\mst22{m_{\;\\widetilde{t}_{2}}}
\def\mst12{m_{\;\\widetilde{t}_{1,2}}}

\def\msb11{m_{\;\\widetilde{b}_{1}}}
\def\msb22{m_{\;\\widetilde{b}_{2}}}
\def\msb12{m_{\;\\widetilde{b}_{1,2}}}

\def\mwidetilde2{\\widetilde{m}^{2}}
\def\lambdawide\widetilde{\\widetilde{\lambda}}
\def\Lambdawide\widetilde{\\widetilde{\Lambda}}

\relax

\begin{document}
\input epsf

\topmargin-2.5cm
%
\begin{titlepage}
\begin{flushright}
CERN-TH/99-117\\
hep--ph/9905242 \\
\end{flushright}
\vskip 0.3in
\begin{center}
{\Large\bf Production of Massive Fermions at Preheating}
\vskip 0.2cm 
{\Large\bf and}
\vskip 0.2cm
{\Large\bf Leptogenesis}

\vskip .5in
{\large\bf G.F. Giudice$^1$}, {\large\bf M. Peloso$^2$}, 
{\large \bf A. Riotto$^1$}, and {\large\bf I. Tkachev$^1$}

\vskip1cm

$^{1}$ {\it CERN Theory Division,
CH-1211 Geneva 23, Switzerland.}

\vskip 0.5cm

$^{2}$ {\it International School for Advanced Studies, via Beirut 
4, I-34014, Trieste, Italy.}

\end{center}
\vskip1.3cm
\begin{center}
{\bf Abstract}
\end{center}
\begin{quote}

We present a complete computation of the inflaton decay into very massive
fermions during preheating. We show that heavy fermions are produced very
efficiently up to masses of order $10^{17}$--$10^{18}$~GeV; the accessible mass
range  is thus even broader than the one for heavy bosons. We apply our
findings to the leptogenesis scenario, proposing a new version of it, in which
the massive right-handed neutrinos, responsible for the generation of the
baryon asymmetry, are produced during preheating. We also discuss other
production mechanisms of  right-handed neutrinos in the early Universe,
identifying the neutrino mass parameters for which the observed baryon
asymmetry is reproduced.

\end{quote}
\vskip1.cm
\begin{flushleft}
May 1999 \\
\end{flushleft}

\end{titlepage}
\setcounter{footnote}{0}
\setcounter{page}{0}
\newpage
%
\baselineskip=20pt
\noindent

\section{Introduction}

It is commonly   believed that the Universe underwent an
early era of cosmological inflation~\cite{lr}. The flatness and the horizon 
problems of the standard big bang
cosmology are indeed elegantly solved if, during the evolution of 
the early
Universe, the energy density happens to be dominated by the vacuum energy 
of a scalar field -- the inflaton -- and comoving scales grow 
quasi-exponentially. 

At the end of inflation the Universe was in a cold, low-entropy state with
few degrees of freedom, very much unlike the present hot, high-entropy
Universe.  
At this stage,
the Universe does not contain any matter and therefore 
it looks perfectly baryon symmetric.
However, considerations
about how the light element abundances were formed when the Universe was about
1 MeV hot lead us to conclude that $n_B/s=(2$--$9)\times 10^{-11}$. Here
$n_B/s$ is
the difference between the number density of
baryons and that of antibaryons, normalized to the
entropy density of the Universe. 

 Until now, several mechanisms for the generation of the baryon ($B$) asymmetry
have been proposed~\cite{rt}.
Grand Unified Theories (GUTs) unify the strong with the electroweak forces
and predict baryon-number violating interactions at tree level. 
In these theories, the out-of-equilibrium
decay of heavy Higgs particles can indeed
explain the observed baryon asymmetry. In the
theory of electroweak baryogenesis, baryon number violation takes place at the
quantum level, caused by unsuppressed baryon-number violating sphaleron
transitions in the hot plasma~\cite{krs}. 

Since $B$ and $L$ -- where $L$ is the lepton number -- are reprocessed
by sphaleron transitions, while the anomaly-free linear  combination $B-L$ is
left unchanged, the baryon asymmetry may be generated 
from a lepton asymmetry~\cite{fy}. 
Indeed,  once the lepton number is produced, thermal scatterings  redistribute
the charges and convert (a fraction of) $L$ into baryon number. In the  
high-temperature phase of the standard model,  the asymmetries of
baryon number $B$ and of $B-L$ are therefore proportional~\cite{kl}:
\be
     B=a(B-L),~~~~a\equiv\left(\frac{8 n_g+4 n_H}{22 n_g+13 n_H}\right),
\label{acoef}
\ee
where $n_H$ is the number of Higgs doublets and $n_g$ is the number of fermion
generations. 

In the standard model as well as in its unified extension
based on the group $SU(5)$, $B-L$ is conserved and no asymmetry in $B-L$ can be
generated. However, adding  massive
right-handed Majorana neutrinos to the standard
model breaks
$B-L$ and the primordial lepton asymmetry may be  generated by their
out-of-equilibrium decay. This simple extension of the standard model
can be embedded into GUTs, as in the case of
$SO(10)$. Heavy right-handed Majorana neutrinos
are the key ingredient to explain the smallness of the light neutrino masses
via the see-saw mechanism~\cite{gelmann}. The presence of neutrino masses and
mixings seems to be the most natural explanation of the    recent reports from
the Super-Kamiokande~\cite{sk} and other~\cite{other} collaborations indicating
the existence of neutrino oscillations. In light of
these considerations, the generation of the baryon  asymmetry through
leptogenesis looks particularly attractive.

The leptogenesis scenario depends crucially on the 
mechanism that was responsible for populating the early Universe with
right-handed neutrinos, and consequently on the
thermal history of the
Universe, and on the fine details of the reheating process after inflation.
One goal of this paper is to discuss several production
mechanisms of heavy right-handed neutrinos in the Universe, to compare them and
to identify the regions of the appropriate parameter space where the production
mechanism is efficient enough to explain the observed baryon asymmetry.

The simplest way to envision the reheating process after inflation
is to assume that the comoving energy
density in the zero mode of the inflaton decays {\it perturbatively} into
ordinary particles, which then scatter to form a thermal
background~\cite{dolgov}. It is usually assumed that the decay width of
this process is the same as the decay width of a free inflaton field. Of
particular interest is a quantity known as the reheat temperature, denoted as
$T_{RH}$. This is calculated by assuming an instantaneous conversion of the
energy density in the inflaton field into radiation when the decay width of the
inflaton $\Gamma_\phi$ is equal to $H$, the expansion rate of the
Universe. This yields 
\be
T_{RH}\simeq \sqrt{ \Gamma_\phi \mpl },
\label{thh}
\ee
where $\mpl$ is the Planck mass.
The commonly-accepted  assumption   is that the heavy right-handed
neutrinos with mass $M_N$ were as abundant as photons at very high
temperatures. This assumption requires not only that $T_{RH}\simgt M_N$, but
also  that the heavy neutrinos are abundantly produced by thermal scatterings
during the reheating stage. This condition, as we will discuss in sect.~3.1,
significantly limits the allowed range of neutrino masses compatible with
leptogenesis.

There might be  one more problem associated with the hypothesis that
$T_{RH}\simgt M_N$ in  the old theory of reheating, and that is
the problem of relic
gravitinos~\cite{gravitino}. If one has to invoke supersymmetry to preserve the
flatness of the inflaton potential, it is mandatory to consider the
cosmological implications of the gravitino -- the spin-3/2 partner
of the graviton which
appears in the  extension of global supersymmetry to supergravity. The slow
gravitino decay rate leads to a cosmological problem because the  decay
products of the gravitino destroy light nuclei by photodissociation and
hadronic showers, 
thus ruining the successful predictions of nucleosynthesis. The requirement that
not too many gravitinos are produced after inflation provides an upper bound
on the reheating temperature $T_{RH}$ of about $10^{8}$--$10^{10}$~GeV,
depending on the value of the gravitino mass~\cite{sarkar}.
In the following, $T_{RH}$ will  be therefore  intended as the largest
temperature allowed after inflation from considerations of the gravitino
problem.

In order to relax the limit on $M_N$ imposed by the gravitino problem,
we will consider the possibility
that the heavy neutrinos are produced directly through the
inflaton decay process. This is kinematically accessible whenever
\begin{equation}
M_{N}< m_\phi,
\end{equation}
where $m_\phi$ is the inflaton mass.
 In the case of chaotic inflation with quadratic potential, the density and
temperature fluctuations observed in the present Universe determine $m_\phi$
and require
$M_{N}$ to be smaller than about $10^{13}$ GeV.

The outlook for leptogenesis   might be brightened even further  with the
realization that reheating may differ significantly from the simple picture
described above~\cite{explosive,KT1,KT2,KT3,KT4}.  In the first stage of
reheating, called {\it preheating}~\cite{explosive}, nonlinear quantum effects
may lead to extremely effective dissipative dynamics and explosive particle
production, even when single particle decay is kinematically forbidden. In this
picture, particles can be produced in a regime of broad parametric resonance,
and it is possible that a significant fraction of the energy stored in the form
of coherent inflaton oscillations at the end of inflation is released after
only a dozen or so oscillation periods of the inflaton. 
The preheating stage occurs because, for some  parameter
ranges, there are new non-perturbative decay channels. 
In the case of bosonic particle production, coherent oscillations of
the inflaton field induce stimulated particle emissions into energy
bands with large occupancy numbers~\cite{explosive}. The modes in these bands
can be understood as Bose condensates, and they behave like classical waves. A
crucial observation for GUT baryogenesis induced by the decay of very heavy
Higgs
bosons is that  particles with masses larger than that of the inflaton may be
easily  produced during preheating~\cite{klr}.   Indeed, for coupling constants
of order unity one would have copious production of  boson particles as heavy
as $10^{15}$ GeV, {\it i.e.} 100 times greater than the inflaton mass.  This
is a major departure from the old constraint of reheating.

In this paper we wish to provide the first complete calculation of the inflaton
decay into heavy fermions during preheating. We will show that fermion
production is extremely efficient, in a mass range even broader than the one
for heavy bosons: fermions may be generated up to masses of order of $10^{18}$
GeV. What distinguishes the production of very massive fermions and bosons in
an oscillating background is the expression for the total mass. For 
bosons, the total mass can never vanish and the production reaches the maximum
when the amplitude of the inflaton goes through zero. For fermions, 
the total mass can
vanish for particular values of the inflaton field, rendering
particle creation much easier. 
 
We will  numerically compute the density of the massive fermions
produced at the resonance stage. This is the crucial parameter 
for leptogenesis, when these heavy fermions are 
identified with the right-handed neutrinos giving rise to the lepton
asymmetry. We want to stress that
the out-of-equilibrium condition is naturally achieved in this
scenario, since the distribution function of the fermionic quanta generated at
the resonance is far from a thermal distribution.  
We will  show that the observed baryon asymmetry 
may be explained by the phenomenon of leptogenesis after preheating, with a
reheating temperature compatible with the gravitino problem.

The paper is organized as follows. In sect. 2 we present our calculation,
with numerical
results as well as some analytical estimates concerning the production of
massive fermions during preheating. In sect. 3 we discuss and compare
the relevant production mechanisms of right-handed neutrinos in the early
Universe, identifying the appropriate neutrino-mass parameters where these
mechanisms are the most efficient, as far as the generation of the baryon
asymmetry is concerned.

\section{Heavy Fermion Production at Preheating}

In this section we describe the basic physics underlying the mechanism of 
heavy fermion  generation during the preheating stage,  
perform the  relevant numerical calculations and present some analytical 
estimates.
In the following we will  focus 
on  the model of chaotic inflation, with a massive 
inflaton $\phi$ having quadratic potential 
$V(\phi)=\frac{1}{2}m_\phi^2\phi^2$. Here  $m_\phi\sim 10^{13}$ GeV is 
fixed by the COBE normalization of  the cosmic microwave background 
anisotropy.

We will suppose that the inflaton field  is coupled 
to a very massive Dirac fermion $X$ with bare mass $m_X$ via  
the Yukawa coupling 
\be
{\cal L}_Y=g\phi\bar{X}X.
\label{Yuky}
\ee 
The total mass of the fermion $X$ is then given by
\be
m(t)=m_X+g \phi(t).
\label{mass}
\ee
When we apply our results to the 
leptogenesis scenario, the fermion $X$ will be identified  
with the lightest of the right-handed Majorana neutrinos 
$N_1$. Although in this section we will restrict our considerations 
to the case of Dirac 
particles, the treatment of Majorana fermions is completely analogous.
Our final results 
regarding 
the abundances of particles (with equal amount of anti-particles
being produced) 
are valid for both 
Dirac and Majorana fermions.

\subsection{The Basic Formalism} 

We start (see {\it e.g.} discussion in ref.~\cite{mmf})
by canonically quantizing the  action of the  massive field $X$ in curved
space with Friedmann-Robertson-Walker metric. 
In the system of coordinates in which the line element is given by   
$ds^2= a^2(\eta)(d\eta^2-  d{\vec x}^2)$, where $a$ is the scale factor of 
the expanding Universe and $\eta$ is the conformal time defined as 
$d\eta=dt/a$, 
the Dirac equation becomes
\begin{equation} 
\label{dirac}
\left( \frac{i}{a} \gamma^\mu \partial_\mu + i  \frac{3}{2} H \gamma^0-m
\right) X = 0.
\end{equation}
Here $H=(a^\prime/a^2)$ is the Hubble rate, the prime denotes derivative
with respect to conformal time, and the $\gamma$-matrices are 
defined in {\it flat} space-time.
By defining $\chi =  a^{-3/2}X$, eq. (\ref{dirac}) can be reduced to 
the more familiar form
\be
\left( i \gamma^\mu \partial_\mu - a \, m \right) \chi = 0.
\label{eqmot}
\ee

Since $a$ is a function of $\eta$, but not of $\vec x$, spatial translations
are symmetries of space-time, and we can separate the variables using the
decomposition
\be
\chi \left( x \right) = \int \frac{d^3k}{(2\pi )^{3/2}} e^{-i\vec{k}\cdot
\vec{x}} \sum_r \left[ u_r(k,\eta)a_r(k) + v_r(k,\eta)b_r^\dagger (-k)\right] ,
\ee
where the summation is over 
spin, and $v_r(k)=C{\bar u}^T_r (-k)$.
We impose the canonical
anticommutation relations on the creation and annihilation operators 
\be 
\left \{ a_r(k),a_s^\dagger (k')\right \} =
\left \{ b_r(k),b_s^\dagger (k')\right \} = \delta_{rs} \delta (\vec{k} -
\vec{k}')
\ee
which, together with the quantization conditions, determine the normalization
of the spinors $u$,
\be
u_r^\dagger (k,\eta) u_s (k,\eta) = v_r^\dagger (k,\eta) v_s (k,\eta)
=\delta_{rs},~~~~~ u_r^\dagger (k,\eta) v_s (k,\eta) =0. \label{confi}
\ee
Equations~(\ref{confi}) are valid at any conformal time, since they
are preserved by the evolution. 

In the representation in which $\gamma^0 =\pmatrix{ \identity &0\cr 
0&-\identity}$ and
with the definition $u\equiv \pmatrix{u_+\cr u_-}$, the equation
of motion~(\ref{eqmot}) can be written as a set of uncoupled second-order
differential equations,
\be
\left[ \frac{d^2}{d\eta^2} +\omega^2 \pm i (a'm+am')\right] u_\pm (k)=0
\label{mot2}
\ee
\be
\omega^2 =k^2+m^2a^2 .
\ee
We can now write the Hamiltonian as
\begin{eqnarray} \label{h}
H \left( \eta \right)=\int d^3 x \chi^\dagger \left( -i \partial_0
\right) \chi &=& \int d^3k\sum_r 
\left\{ E_k
\left( \eta \right) \left[ a_r^\dagger (k) a_r(k)- b_r(k) b_r^\dagger (k)
\right]
\right. \nonumber\\
&+& \left.  \, F_k \left( \eta \right) b_r(-k)a_r(k)
+  F_k^*
\left( \eta \right) a_r^\dagger (k) b_r^\dagger (-k)
\right\} .
\end{eqnarray}
By using the equations of motion, we find
\bea
&&E_k =2k~ {\rm Re} (u_+^*u_-) +am
\left( 1-2u_+^*u_+\right) , \\
&&F_k = k \left(u_+^2 -u_-^2\right) + 2am~
u_+u_- , \\
&&E_k^2 +|F_k|^2 =\omega^2.
\eea
Here we have chosen the momentum $k$ along the third axis, 
and selected the gamma-matrix representation in which
$\gamma^3=\pmatrix{ 0&\identity \cr -\identity &0}$.

In order to give a ``quasi-particle" interpretation, we diagonalize 
the Hamiltonian
in eq.~(\ref{h})
with a time-dependent Bogolyubov canonical transformation, and define the
new creation and annihilation operators
\begin{eqnarray}
&&{\hat a} (k,\eta)= \alpha(k,\eta)a(k)+\beta(k,\eta) b^\dagger (-k)
\\
&&{\hat b}^\dagger (k,\eta)= -\beta^*(k,\eta)a(k)+\alpha^*(k,\eta) 
b^\dagger (-k).
\end{eqnarray}
Imposing canonical anticommutation relations on the operators ${\hat a}$
and ${\hat b}$, we find $|\alpha|^2+|\beta|^2=1$. For
\be 
\frac{\alpha}{\beta}=\frac{E_k+\omega}{F^*_k},~~~~|\beta|^2=
\frac{|F_k|^2}{2\omega (\omega +E_k)}=\frac{\omega -E_k}{2\omega},
\ee
the normal-ordered Hamiltonian in terms of 
the ``quasi-particle" operators is diagonal,
\be
H \left( \eta \right) = \int d^3k\sum_r 
\omega \left( \eta \right) \left[ 
{\hat a}_r^\dagger (k) {\hat a}_r(k)+ {\hat b}^\dagger_r(k) {\hat b}_r(k)
\right] .
\ee

Next, we define a ``quasi-particle" vacuum, such that $\hat a |0_\eta\rangle
=\hat b |0_\eta\rangle =0$. The total number of produced particles up to
time $\eta$ (equal to the number of produced antiparticles) is given
by the vacuum expectation value of the particle number operator $N$,
\be
n(\eta)=\langle 0_\eta |N|0_\eta\rangle=\frac{1}{\pi^2 a^3(\eta)}
\int_0^\infty dk~ k^2~ |\beta |^2.
\ee
The density of produced particles is then computed 
by integrating the equations of motion~(\ref{mot2}) with an initial
condition at time $\eta =0$ given by\footnote{In practice, one has to
solve only the evolution equation for $u_+$, since $u_-$ is determined
by the equation $u_-=(amu_+-iu_+^\prime )/k$.}
\be
u_\pm (0)=\sqrt{\frac{1}{2}\left( 1\mp \frac{ma}{\omega}\right)},~~~~
u^{\prime}_\pm (0)=iku_\mp (0) \mp iamu_\pm (0) .
\ee 
This boundary condition 
corresponds to $E_k =\omega$, $F_k=0$ at $\eta=0$ or, in other words, to
an initial vanishing particle density.

\subsection{Numerical Results}

The equations of motion~(\ref{mot2}) describe oscillators with time-varying
complex frequency. If $m$ is constant, the time dependence enters only
through the scale factor $a$. 
The corresponding gravitational 
creation of heavy fermions was studied in ref.~\cite{KT99}.
However, of particular interest is
the case in which  $m=m(t)$ is a periodic function of time.
This is realized when the scalar
field $\phi$, coupled to $X$ as in eq. (\ref{Yuky}) is homogeneous
and oscillates in time with frequency  $V^{\prime\prime}(\phi)$.
It is useful
to write the equations of motion in terms of
dimensionless variables.
We introduce a dimensionless time $\tau \equiv m_\phi\eta $,
as well as a dimensionless field $\varphi \equiv \phi /\phi(0)$,
so that the scalar field is normalized by the condition
$\varphi (0) =1 $. We define $\phi(0)$ as the value of the inflaton field
at the moment when its oscillations begin,
$\phi(0) \simeq 0.28 \mpl$ \cite{KT2}.
With these redefinitions, the equation of motion for the 
background field $\varphi$ 
does not contain any parameter, while  the fermion mass is measured in
units of $m_\phi$ and the strength of the fermion coupling to the external 
background is determined by the dimensionless combination
$g\phi(0)/m_\phi$. For the sake of correspondence with the bosonic case, we
introduce the parameter 
\be
q \equiv g^2\phi^2(0)/4m_\phi^2 \, .
\label{q}
\ee

The production of fermions by an external oscillating background 
in Minkowski space-time was studied in refs.~\cite{gmm,BHP98,GK98}.
The production of massless fermions in a  $\lambda \phi^4$ inflaton model,
studied in ref.~\cite{GK98}, can be reformulated  into particle production
in an  expanding Universe by
a simple  conformal transformation. Moreover, only the case of moderate $q$
($q < 100$) has been previously considered. 
The parametric production 
of massive fermions 
in an expanding Universe has never been analyzed before. By analogy with the 
bosonic case~\cite{explosive,KT2,KT3}, we expect that an
efficient production of very massive fermions will require
a very large value of $q$.

\begin{figure}
\centerline{\leavevmode\epsfysize=8cm \epsfbox{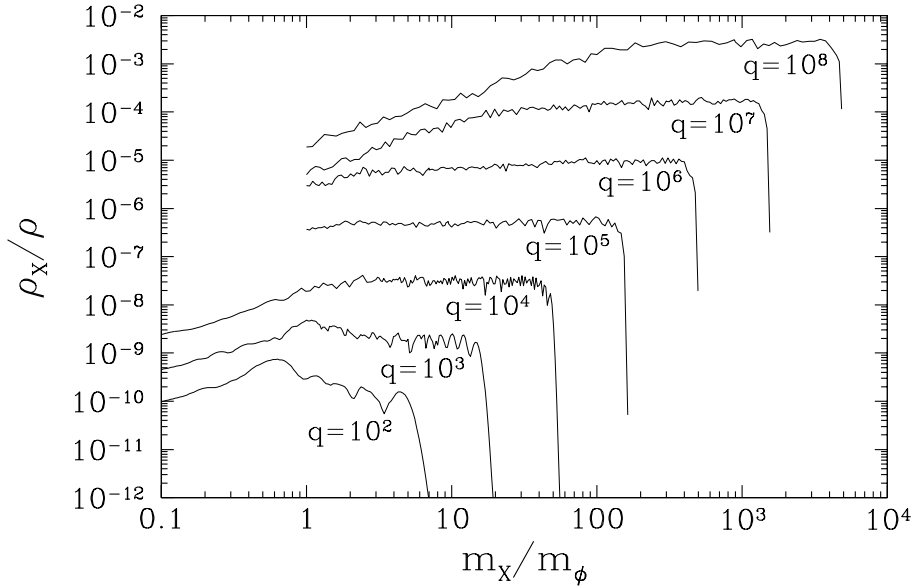}}
\vspace{10pt}
\caption{The fraction of the energy density of produced fermions 
with respect to the total energy density,
as a function of the fermion mass $m_X$ in units of the inflaton mass,
for various 
values of $q$.}
\label{fig:mscan}
\end{figure}

The result of our numerical integration is best summarized
in fig.~\ref{fig:mscan}. For different values of 
the fermion mass $m_X$
and of
the $q$ parameter,
fig.~\ref{fig:mscan} shows $\rho_X/\rho$, 
the fraction of
the inflaton energy density  $\rho$  which ends up in the 
fermionic energy density 
$\rho_X$. In our simulation, we
have neglected the back  reaction of the produced fermions in the evolution
of the inflaton field.

\begin{figure}
\centerline{\leavevmode\epsfysize=8cm \epsfbox{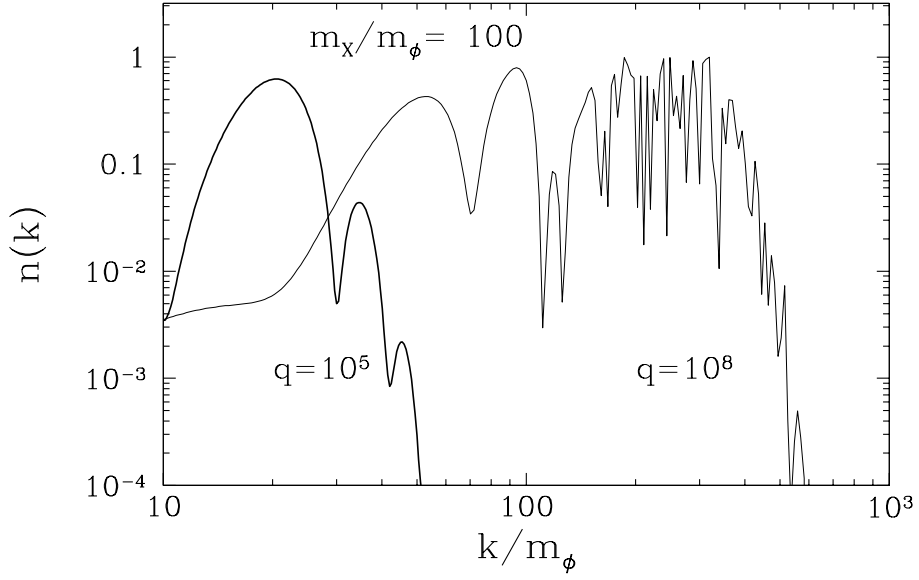}}
\vspace{10pt}
\caption{The final phase-space density of produced particles for two values
of the parameter $q$. The $X$-fermions are taken to be  
100 times heavier than the 
inflaton. At $q=10^5$ the freeze-out 
of the particle production  was reached after the first inflaton oscillation,
while for $q=10^8$ it required twenty oscillations.} 
\label{fig:pws1}
\end{figure}

Fermion production is 
efficient up to a time at which 
ratio $\rho_X/\rho$  freezes out.  For fixed $q$, the larger $m_X$ is,
the earlier this freeze-out  occurs. In particular, near the 
cut-off of 
$\rho_X/\rho$ at large $m_X$, the production ceases just after a few
oscillations of the inflaton field. 
In fig.~\ref{fig:pws1} we have plotted the  final phase-space density
of produced fermions in comoving volume for $m_X = 100~ m_\phi$,  
$q=10^5$ and $q=10^8$. 
For $q=10^8$ this distribution is reached after twenty inflaton oscillations,
while for $q=10^5$ it is reached just after the 
first inflaton oscillation.
In fig.~\ref{fig:mscan}, the curves for $q=10^7$ and $q=10^8$ correspond
to an 
evolution up to 20 inflaton oscillations. For $q=10^8$ 
and $m_X<100~m_\phi$, the freeze-out has not been reached in 20
oscillations and $\rho_X/\rho$ would have grown further, 
had we  integrated for longer times.
This explains the slope of the curves in fig.~\ref{fig:mscan} at small
$m_X$ for $q>10^6$.
In general, if we integrate up to the freeze-out,
we find that the ratio  $\rho_X/\rho$ is almost constant
with $m_X$ up to a cut-off value $(m_X)_{\rm max}$ much larger than $m_\phi$.
In this regime,
the number density of fermions is inversely proportional to 
$m_X$. 

\begin{figure}
\centerline{\leavevmode\epsfysize=8cm \epsfbox{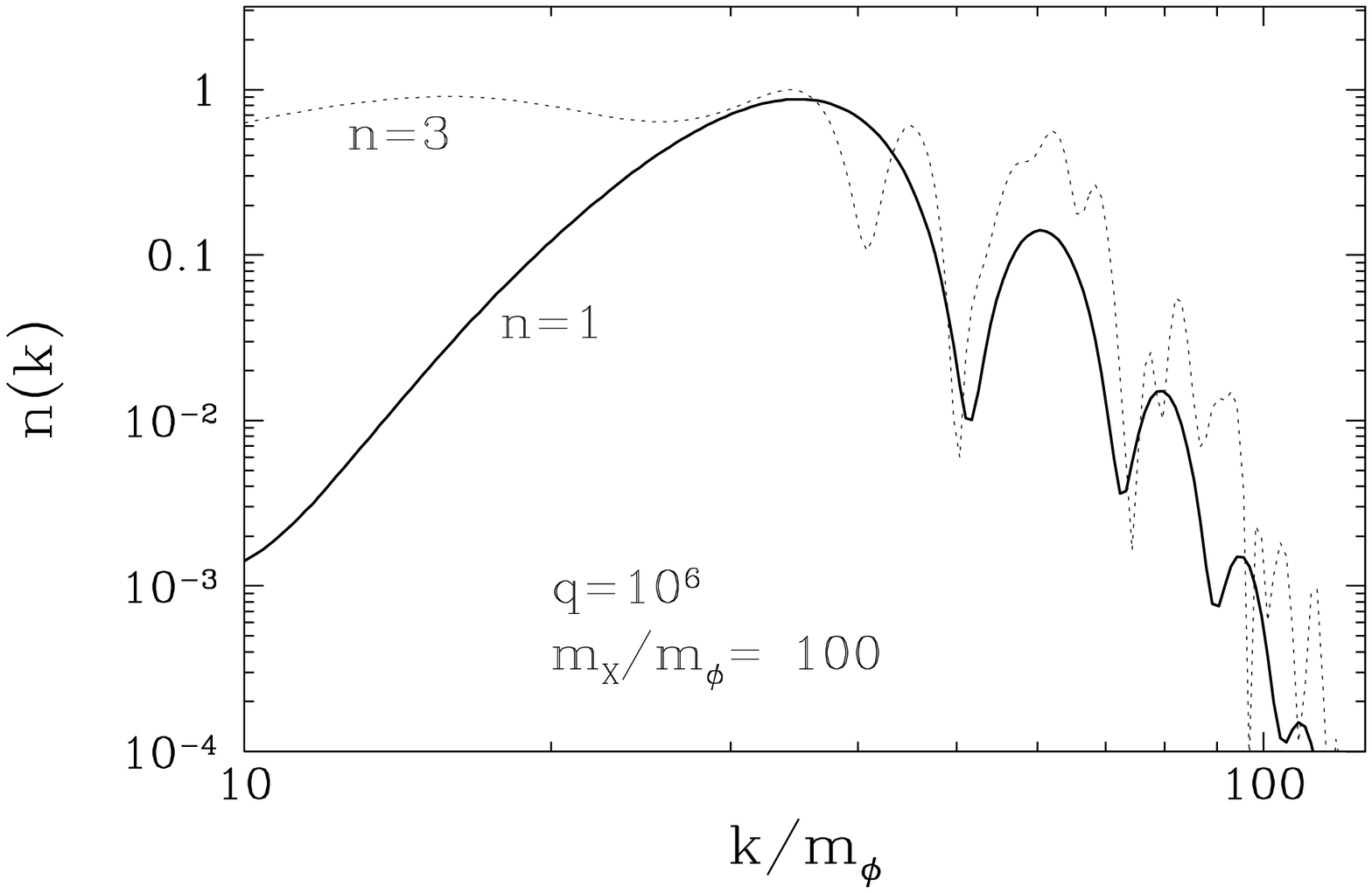}}
\vspace{10pt}
\caption{The phase-space density of produced particles after $n=1$ and $n=3$
 inflaton oscillations for $q=10^6$ and for
$X$-fermions 100 times heavier than the inflaton. The 
distribution at $n=3$ coincides with the final distribution.} 
\label{fig:pws2}
\end{figure}

\begin{figure}
\centerline{\leavevmode\epsfysize=8cm \epsfbox{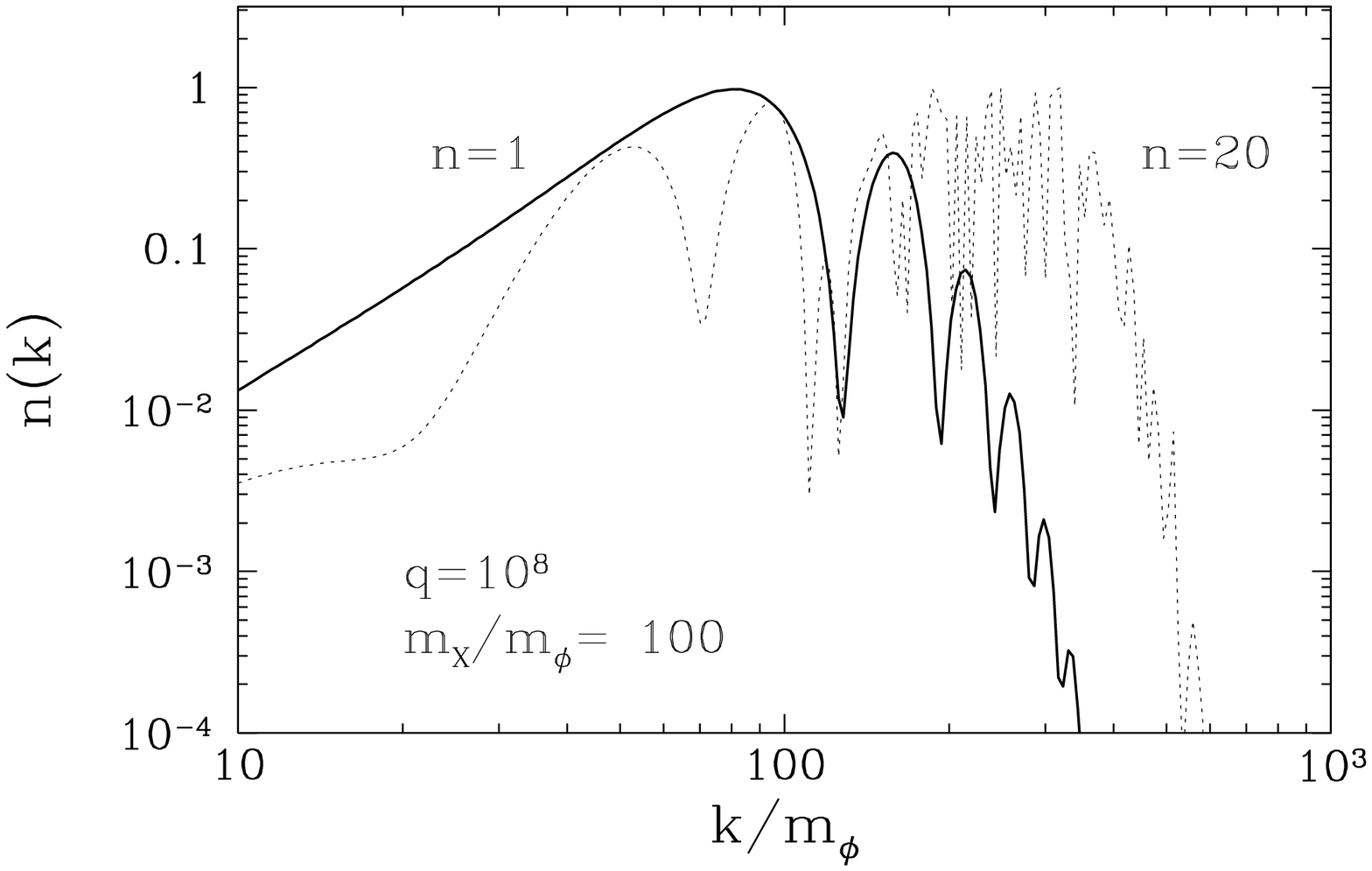}}
\vspace{10pt}
\caption{The phase-space density of produced particles after $n=1$ and $n=20$
inflaton oscillations for $q=10^8$ and for $X$-fermions 
100 times heavier than the inflaton. The distribution at $n=20$ coincides 
with the final distribution.} 
\label{fig:pws3}
\end{figure}

The time evolution of the phase-space
density of produced fermions in the cases in which more  than
one inflaton oscillation is required to reach the final distribution is shown
in figs.~\ref{fig:pws2} and ~\ref{fig:pws3}. For the  parameters
shown in fig.~\ref{fig:pws2}, 
the final distribution was reached after three inflaton oscillations,
while in the case of fig.~\ref{fig:pws3} the $X$-particle production continues 
up to the twentieth oscillation.

Particle production  at large $q$ and large $m_X$ is similar to
the ``instant preheating'' phenomenon discussed in ref.~\cite{instant}. 
However, contrary to what is discussed in ref.~\cite{instant}, where
 the production of heavy fermions was possible only  through the 
coupling of the fermion field to other  bosonic degrees of freedom  
produced by  preheating, in our case the instant preheating takes place 
by coupling the fermion directly  to the inflaton field, and therefore this
realization  is much less model-dependent. 
 
Let us summarize our  numerical results  in a form suitable for 
the analytical estimates we will perform in sect.~2.3.
Because of the Fermi-statistics, $n(k) \equiv |\beta |^2 \le 1$. The
maximum value of $n(k) \sim 1$ is reached rapidly at some $k = k_{\rm max}$.
With time, if particle creation is still efficient, $k_{\rm max}$ 
grows at each oscillation.
We can estimate the number density of the produced fermions as
$n = \int d^3k n(k) \propto k_{\rm max}^3$. {}From our numerical
results we observe that, at fixed $m_X$, the 
freeze-out value of $k_{\rm max}$ scales as
$k_{\rm max} \propto q^\alpha$, with $\alpha$ slightly larger than
$1/3$; see figs.~\ref{fig:mscan} and ~\ref{fig:pws1}.
On the other hand,  $k_{\rm max}$ has to scale as 
$ m_X^{-1/3}$ to explain why $\rho_X/\rho$ is nearly independent of
$m_X$, at constant $q$.
We also find numerically that the cut-off value $(m_X)_{{\rm max}}$, {\it i.e.}
 the maximum mass
of produced fermions, is about  $(m_X)_{\rm max} \simeq q^{1/2}m_\phi/2$.

\subsection{Analytical Estimates}

The goal of this section is to provide  some  analytical estimates that can
help us to understand   the  numerical results presented above. 

Compared with the bosonic case,
the novel feature of the fermionic
production  is that  much heavier particles can be generated  rather
efficiently during the oscillations of the inflaton field, see
fig.~\ref{fig:mscan}. 
The key difference between the Bose and Fermi cases resides  in the
expression for the effective particle mass in  an external
oscillating background. In the Fermi case it is given by eq. (\ref{mass}),
$m(t)=m_X+g \phi(t)$, while in the Bose case we have
$m^2(t)=m_X^2+g^2\phi^2(t)$. In both cases, the production of  particles
with masses larger than the inflaton mass
({\it i.e.} the   frequency of the inflaton
oscillations) requires large $q$. This, in turn, means that $m(t)$ is typically
large, and creation of $X$-particles is impossible  at  all times,
except for  very short periods when $m(t)$ fluctuates around its minimum value. 
In the bosonic case $m(t)$ can never 
be smaller than $m_X$. However, in the fermionic case,
$m(t)$ can vanish as long as the amplitude of the inflaton field is
large enough, $|\phi | > m_X/g$. 

In the expanding Universe the  amplitude $\phi_0$ of the oscillating 
inflaton field decreases with time. It is reasonable
to assume that the sharp cut-off in the particle production at large $m_X$ that
we  observe  in fig.~\ref{fig:mscan} corresponds to a situation in which
the condition $m(t)=0$ cannot be satified even during the first oscillation of
the inflaton field. To verify this assumption,   let us write $m(t)$, with
the help of eq.~(\ref{q}), as
\be 
m = m_X + 2\sqrt{q}m_\phi
\frac{\phi}{\phi(0)} \simeq m_X + \frac{\sqrt{q}m_\phi}{\pi N}
\,\mbox{cos}(2\pi N)\, .
\label{mass2} 
\ee 
Here we have  taken  into
account that $\phi\propto t^{-1}$, when the energy density of the Universe is 
dominated by the oscillating inflaton field, and we have  denoted with $N$ the
number of oscillations of the inflaton field $N = m_\phi t / (2\,\pi)$. 

At large $m_X$,  the minimum of $m$  is reached around $N = 1/2$. Therefore,
if 
\begin{equation} 
\label{co} m_X> m_\phi\frac{2}{\pi}\:\sqrt{q} , 
\end{equation} 
the
total mass $m$ never vanishes\footnote{We are considering here the case in 
which $g \phi(0)$ and $m_X$ have the same sign. If these two terms
have a relative minus sign,
particle production can be extended to even larger values of $m_X$.}. 
The value of $(m_X)_{\rm max}$ given in
eq.~(\ref{co}) is already in good agreement with our numerical results.
However,
eq.~(\ref{mass2}) cannot be trusted at small
$N$. Actually, our numerical integration shows that the minimum of the inflaton
amplitude
$\phi_0$  during the first
oscillation is given by
\begin{equation} 
\frac{\phi_0}{\phi(0)} \simeq - 0.25. 
\end{equation}  
This means that the
cut-off value of the mass is at 
\begin{equation} 
(m_X)_{{\rm max}} \simeq
m_\phi \frac{\sqrt{q}}{2} , 
\end{equation}   
in  perfect agreement with the results
presented in fig.~\ref{fig:mscan}. 

If $m_X<(m_X)_{{\rm max}}$, $m(t)$ can vanish  more than once during the
inflaton oscillations. In particular, in the case in which  $m_X \gg m_\phi$,
one can suppose   that the production of particles continues until the
amplitude of the  oscillations drops below the critical value 
\be
{(\phi_0)_{{\rm crit}}} =\frac{m_X}{g}=
\frac{m_X}{2m_\phi \sqrt{q}}\,{\phi(0)}. 
\label{frees-out} 
\ee 
At later times,
when $\phi_0< (\phi_0)_{{\rm crit}}$, the ratio $\rho_X/\rho$ is frozen,
since particle production has stopped. Let us now estimate analytically  the
final
value of  $\rho_X/\rho$. During the
inflaton oscillations, the Fermi distribution function is rapidly saturated 
up to some maximum value of the momentum $k$, {\it i.e} $|\beta(k)|^2\simeq  1$
for $k\simlt k_{{\rm max}}$ and it is zero otherwise. The value of $k_{{\rm
max}}$ increases at each inflaton oscillation until particle production
stops. Therefore, what is relevant for the determination of the final value of
$\rho_X/\rho$ is $k_{\rm max}$ at the freeze-out, {\it
i.e.} when eq.~(\ref{frees-out}) is satisfied and particles are no
longer generated. At this moment, 
\be 
\rho_X \simeq \frac{m_X p_{\rm
max}^3}{3\pi^2} \, , \label{rhoX} 
\ee 
where $p_{\rm max}$ is the maximum {\it physical}
momentum at freeze-out, $p_{{\rm max}}=k_{{\rm max}}/a$.

We estimate $p_{{\rm max}}$ as the maximum value of the physical momentum 
above which the evolution of the vacuum mode function is adiabatic. 
We   choose as  a trial mode function the following 
expression
\be 
u_{+} (\eta) =
\sqrt{\frac{1}{2}\left(1 - \frac{m_{\rm eff}}{\omega} \right)}
\times \exp\left({-i\int^\eta
\omega\: d\eta'}\right), \label{vac} 
\ee 
where $m_{{\rm eff}}=ma$. Notice that this solution 
corresponds to a vacuum, since it gives $\beta(k) =0$.
This trial
function solves to a   good approximation the equation of motion if the
following condition holds 
\be 
\frac{1}{2}\left(1+\frac{m_{\rm
eff}}{\omega}\right) \left| \frac{m_{\rm eff}''}{\omega} +
\frac{1}{2}\left(\frac{m_{\rm eff}'}{\omega}\right)^2 \left(1-\frac{5m_{\rm
eff}}{\omega}\right) \right| \ll \omega^2, \label{adiabatic} 
\ee 
where we have used the property $\omega'= m_{\rm eff}m_{\rm eff}'/\omega$.
As we have
explained above, for large $q$, particle production takes place 
during short intervals when $m_{\rm eff}\simeq 0$, or
$\cos (m_\phi t) \simeq - m_X /g\phi_0$, neglecting the  expansion of 
the Universe during these short  periods of particle creation. 
Under these circumstances, the condition
(\ref{adiabatic}) reduces to 
\begin{equation} 
\left|2pm_X + g^2\phi_0^2 - m_X^2\right|\, m_\phi^2 \ll 4\: p^4 \,\, ,
\label{adiabatic2}
\end{equation}  
where we have used $\omega\simeq k=pa$.
As a result, at the end of the particle production epoch, see  eq. 
(\ref{frees-out}), we get
\be 
p_{\rm max}^3 \simeq  \frac{1}{2} m_X m_\phi^2. 
\label{pmax}
\ee  
To compare with the numerical results of sect.~2.2, it is convenient to
express eq.~(\ref{pmax})
in terms of the comoving momentum, 
\be
k_{\rm max} = \left(\frac{2m_\phi^4}{m_X} q\right)^{1/3}.
\ee
This reproduces the approximate scaling law $k_{\rm max}\propto q^{1/3}
m_X^{-1/3}$ observed in the numerical results of sect.~2.2.
Notice that,
at  earlier epochs when $g^2\phi_0^2 \gg m_X^2$, the value of $p_{\rm max}$ is
larger, but the corresponding  fraction of $X$-particles produced at this stage
 becomes subdominant (because it is red-shifted) with respect to the fraction
of particles generated right before the freeze-out. In terms of the
parameter $q_{\rm eff}(t) \equiv g^2 \phi^2(t)/4m_\phi^2$, we obtain
that at early times $p_{\rm max} \propto q_{\rm eff}^{1/4}$, while at the end
of the particle production epoch $p_{\rm max} \propto q_{\rm
eff}^{1/6}$. 

Our  result   for  $p_{\rm max}$ amd $(m_X)_{\rm max}$ and their relative 
scaling with $q$ for the fermions  produced during preheating 
 are   different from what 
argued in ref.~\cite{GK98} and  from the behaviour of the same 
quantities encountered when dealing with boson particle production
during  preheating. In particular, for bosons, both 
$p_{\rm max}$ and $(m_X)_{\rm max}$ scale like
$ q_{\rm eff}^{1/4}$~\cite{explosive,KT2}.

We can  now use eqs.~(\ref{rhoX}) and (\ref{pmax}) to find that $\rho_X
\simeq m_X^2m_\phi^2/6\pi^2$. The inflaton energy density at freeze-out is
$\rho = m_\phi^2 \phi^2/2 = \phi^2(0)m_X^2/8q$, and therefore the energy
density fraction is
\be 
\frac{\rho_X}{\rho} \simeq
\frac{4}{3\pi^2}\frac{m_\phi^2}{\phi^2(0)} q 
= \frac{1}{3\pi^2} \, g^2. 
\label{frac1} 
\ee 
This expression describes quite well the behaviour observed in
fig.~\ref{fig:mscan} for a large range of $q$. Equation~(\ref{frac1})
does not depend upon
$m_X$, it is approximately proportional to $q$ and it gives a  reasonable
estimate for the overall magnitude. Indeed, for  $\phi(0) \simeq 0.28 \mpl$
\cite{KT2} and $m_\phi /\mpl \simeq 10^{-6}$, eq. (\ref{frac1}) reduces to
${\rho_X}/{\rho} \simeq 2 \times 10^{-12} q$, in good agreement with our
numerical results. At very large values of  $q$ the  expression (\ref{frac1}) 
underestimates the ratio $\rho_X/\rho$; for instance at $q=10^8$, it
gives  a value of $\rho_X/\rho$ smaller than the numerical result by
about one  order of magnitude.

In the next section, we will apply our results on fermionic preheating  
to the case of heavy right-handed neutrinos and  leptogenesis. We want  
to emphasize, however, that the results obtained in this section may be  
relevant in other cosmological
contexts, such as the generation of superheavy dark  
matter after inflation~\cite{shdm}.

\section{Leptogenesis}

The Lagrangian terms relevant for leptogenesis describe the interactions
between the massive right-handed
neutrinos $N$, the lepton doublet $\ell_L$, and the Higgs doublet $H$,
\be
{\cal L}=-\bar N h_\nu H \ell_L -\frac{1}{2} {\bar N}^c M N +{\rm h.c.}
\ee
Here the Yukawa couplings $h_\nu$
and the Majorana mass $M$ are $3\times 3$ matrices and
generation indices are understood. We choose to work
in a field basis in which $M$ is diagonal with real and positive 
eigenvalues ordered
increasingly ($M_1<M_2<M_3$). The 
mass matrix of the light, nearly left-handed, neutrinos is given by
\be
m_\nu = -h_\nu^TM^{-1}h_\nu \langle H \rangle^2 .
\label{neumas}
\ee

The decays of the heavy neutrinos $N$ into leptons and Higgs 
bosons violate lepton number 
\begin{eqnarray}
N &\rightarrow & \bar H \ell,\nonumber\\
N &\rightarrow & H\bar \ell.
\end{eqnarray}
The interference between the tree-level decay amplitude and the absorptive 
part of the one-loop diagram can lead to a lepton  asymmetry of the right order 
of magnitude to explain the observed baryon asymmetry, as has been 
extensively discussed in the literature~\cite{fy,luty,extra,plum} (for reviews,
see ref.~\cite{revlep}).
The interference with the one-loop vertex amplitude yields a CP-violating
decay asymmetry for $N_1$ equal to 
\be
\epsilon_V = \frac{1}{8\pi \left(h_\nu h_\nu^{\dag}\right)_{11}}\sum_{j=2,3}
  {\rm Im}\left[\left(h_\nu h_\nu^{\dag}\right)_{1j}\right]^2\:f
  \left({M_j^2\over M_1^2}\right),
\ee
\be
   f(x)=\sqrt{x}\left[1-(1+x)\ln\left(\frac{1+x}{x}
  \right)\right] .
\ee
The absorptive part of the one-loop self-energy gives a contribution to the 
$N_1$ asymmetry which, in the case of only two-generation mixing, is  given by
\be
\epsilon_S = \frac{{\rm Im}\left[\left(h_\nu h_\nu^{\dag}\right)_{1j}\right]^2}
{\left(h_\nu h_\nu^{\dag}\right)_{11}\left(h_\nu h_\nu^{\dag}\right)_{22}}
\left[ \frac{(M_1^2-M_2^2)M_1\Gamma_{N_2}}{(M_1^2-M_2^2)^2 + 
M_1^2\Gamma_{N_2}^2} \right] .
\ee
Here $\Gamma_{N_i}$ is the total decay rate of the right-handed neutrino $N_i$,
\be
\Gamma_{N_i}=\frac{\left(h_\nu h_\nu^{\dag}\right)_{ii}}{8\pi}M_i .
\label{width}
\ee
The CP-violating asymmetry $\epsilon_S$ is
enhanced when the mass difference between two heavy right-handed  neutrinos 
is small, although not smaller than the decay width.

The total CP asymmetry $\epsilon$ has an involved dependence on the complete
structure of the neutrino matrices $h_\nu$ and $M$. However, let us 
assume that the lepton asymmetry is generated only at the decay of the
lightest right-handed neutrino $N_1$. This hypothesis is satisfied if the
$N_1$ interactions are in equilibrium at the time of the $N_{2,3}$ decay 
(erasing any produced asymmetry),
or if $N_{2,3}$ are too heavy to be produced after inflation. As will be made
clear in the following, this is a very plausible working assumption. In this
case, the dynamics of leptogenesis can be described in terms of only 3
parameters~\footnote{The other neutrino mass parameters come into
play only for very large values of $m_i\equiv \left(h_\nu 
h_\nu^{\dag}\right)_{ii}\langle H\rangle^2 /M_i$ ($i=2,3$), when
the lepton-number violating interactions mediated by $N_2$ or $N_3$ can
partially erase the lepton asymmetry.}
: $\epsilon$, $M_1$, and
\be
m_1\equiv \left(h_\nu h_\nu^{\dag}\right)_{11}\langle H\rangle^2
/M_1.
\ee
The parameter $m_1$, which determines the relevant interactions of $N_1$,
coincides with the light neutrino mass $m_{\nu_1}$ only in the limit
of small mixing angles, see eq.~(\ref{neumas}).

Here we will be mostly concerned with the production mechanisms of
$N_1$ in the early cosmology. We will discuss a variety of these mechanisms and
identify, in the different ranges of $M_1$ and $m_1$, the size of
$\epsilon$ required to generate the appropriate baryon asymmetry,
$n_B/s\sim (2$--$9)\times 10^{-11}$. 
Our results can be used to check if specific
particle-physics models for neutrino mass matrices are compatible with the
various leptogenesis mechanisms.

\subsection{Thermal Production}
\label{secter}

We start by considering the case in which the right-handed neutrino $N_1$
reaches thermal equilibrium by scattering with the bath after the inflaton
decay. The amount of lepton asymmetry generated by the $N$ decay can then
be computed by integrating the appropriate Boltzmann equations~\cite{luty,plum}.

A measure of the
efficiency for producing the asymmetry is given by the ratio $K$
of the thermal average of the $N_1$ decay rate and the Hubble parameter
at the temperature $T=M_1$,
\be
K\equiv \left. \frac{\Gamma_{N_1}}{2H}\right|_{T=M_1} =
\frac{m_1}{2\times 10^{-3}~{\rm eV}} .
\ee
Here we have expressed the $N_1$ decay width, see eq.~(\ref{width}), in terms
of the parameters $m_1$ and $M_1$ as
\be
\Gamma_{N_1}=\frac{G_F}{2\sqrt{2}\pi}m_1M_1^2 .
\label{decad}
\ee

For $m_1 \simlt 2\times 10^{-3}$~eV, $K$ is less than unity and the decay
process is out of equilibrium when $N_1$ becomes non-relativistic.
Under these conditions, the leptogenesis becomes very efficient. Indeed,
the produced baryon asymmetry approaches
its theoretical maximum value obtained
by assuming that each $N_1$ in thermal equilibrium eventually generates
$\epsilon$ baryons,
\be
\left(\frac{n_B}{s}\right)_{\rm max}=
\frac{135~\zeta (3) ~a}{4\pi^4g_*}\epsilon = 1\times 10^{-3} 
\epsilon .
\label{max}
\ee
Here $a$ is defined in eq.~(\ref{acoef}) and $g_*$ counts the number of degrees
of freedom (for the standard model particle content, $a=28/79$ and $g_*=
427/4$).

For very small $m_1$, $K\ll 1$ and $N_1$ decouples when it is still 
relativistic. At temperatures $T$ below $M_1$, the $N_1$ 
contribution to the energy
density red-shifts like matter and therefore $\rho_{N_1}/ \rho_{total}
=(7M_1)/(4g_*T)$. Eventually $N_1$ matter-dominates
the Universe at a temperature
\be
T_{dom} =\frac{7~M_1}{4~g_*}\simeq 2\times 10^{-2} M_1 .
\ee
However this can happen only if $N_1$ does not decay beforehand. Since the
decay temperature is 
\be
T_*=0.8g_*^{-1/4}\sqrt{\Gamma_{N_1}\mpl}=
\left( \frac{m_1}{10^{-6}~{\rm eV}}\right)^{1/2} \left(
\frac{M_1}{10^{10}~{\rm GeV}}\right) ~
3\times 10^{8}~{\rm GeV} ,
\label{tstar}
\ee
neutrino matter-domination occurs when
\be
m_1< 4\times 10^{-7}~{\rm eV}.
\ee
Under these conditions,
the bulk of the energy of the Universe is stored
in the non-relativistic $N_1$.  At the time of decay,
such energy density  
is converted into  relativistic degrees of freedom whose
 temperature coincides with  $T_*$ given in eq. (\ref{tstar}),
\be
\rho_{N_1}=M_1n_{N_1}= \frac{\pi^2}{30}g_*T_*^4.
\ee
This yields the following baryon asymmetry
\be
\frac{n_B}{s}=\epsilon a \frac{n_{N_1}}{s}=\epsilon a \frac{3T_*}{4M_1}=
\left( \frac{m_1}{10^{-6}~{\rm eV}}\right)^{1/2}8 \times 10^{-3}\epsilon  .
\label{baune}
\ee
In order not to reintroduce the cosmological gravitino problem, discussed in the
introduction, one 
has to require that the temperature $T_*$ after the
right-handed neutrino decay is less than the maximum value allowed 
 $T_{RH}$. This   implies
\be
m_1< \left( \frac{T_{RH}}{M_1} \right)^2 10^{-3}~{\rm eV}.
\ee
Therefore, the  leptogenesis process is very efficient also when $T_{RH}$ 
is low enough to suppress gravitino production.

For $K \gg 1$, the departure from thermal equilibrium is reduced,
and leptogenesis is less efficient. 
Larger values of $\epsilon$ are now required. However, for
$m_1\sim 10^{-2}$~eV, values of $\epsilon$ of about
$10^{-5}$ are sufficient to generate the appropriate baryon asymmetry.
Realistic neutrino mass matrices can
comfortably reproduce such values of $\epsilon$.
Notice that the prediction for the baryon asymmetry depends weakly
on $M_1$, as long as $m_1$ is not too large~\cite{plum}. 
This is because both the production and decay thermal rates
of $N_1$ at $T=M_1$
depend only on $m_1$, while the
$M_1$ dependence arises from lepton-violating
$H$--$\ell$ scattering.

Let us now turn to discuss how primordial thermal equilibrium of $N_1$ can be
achieved. The first necessary condition is 
\be
M_1<T_{RH} .
\label{cond1}
\ee
This can be quite constraining, especially in view of the bound derived from
the disruptive gravitino effects on nucleosynthesis discussed mentioned in the
introduction.
When the inequality~(\ref{cond1}) is not satisfied,
one expects  the number density of the
$N_1$-particles generated during the reheating stage to be quite small,
making 
this case marginal, as far as the generation of baryon number is 
concerned. We would only like to mention here that such a number density    
depends upon the fine details of the  dynamics of the reheating stage itself.
In particular, the reheat temperature $T_{RH}$
is not the maximum temperature obtained after inflation; the
maximum temperature is, in fact, much larger than $T_{RH}$~\cite{ckr3}. 
As a result, the abundance of massive particles may be suppressed
only by powers of the mass over the temperature, and not exponentially.

A second condition for $N_1$ thermalization 
is derived from the requirement that inverse decay or
production processes of the kind $\bar \ell q_{(3)}\to N_1 t$ (mediated
by Higgs-boson exchange) are in thermal equilibrium before $N_1$ becomes
non-relativistic. This implies
\be
m_1\simgt 10^{-3}~{\rm eV}.
\label{limeq}
\ee
This condition excludes the possibility of the most efficient leptogenesis
with $K<1$. However, even if $m_1$ is somewhat smaller than the value indicated
by eq.~(\ref{limeq}), a sufficient number of $N_1$ can be produced.
Indeed, for $m_1\simgt 10^{-5}~{\rm eV}$, values of $\epsilon \simgt 10^{-5}$
can give
rise to the observed baryon asymmetry. In the case of supersymmetric models,
the constraint can be even less stringent~\cite{plum2}, and values
$\epsilon \simgt 10^{-5}$ are sufficient for $m_1\simgt 10^{-6}~{\rm eV}$.

The constraint on $m_1$ from thermalization
can be evaded if new interactions, different
from the ordinary Yukawa forces, bring $N_1$ in thermal equilibrium at high
temperatures. For instance,
one could use the extra $U(1)$ gauge interactions included
in $SO(10)$ GUTs. These interactions can produce a thermal population of
$N_1$ if, at $T=T_{RH}$,
\be
\Gamma (\bar f f\to Z'\to NN)=\frac{169 ~\alpha_{GUT}^2~T^5}{3\pi ~M_{Z'}^4}>H.
\ee
This requires that the mass of the extra gauge boson $M_{Z'}$ should be close
to $T_{RH}$ and significantly lower than the GUT scale,
\be
M_{Z'}< \left( \frac{T_{RH}}{10^{10}~{\rm GeV}}\right)^{3/4}  
4\times 10^{11}~{\rm GeV}.
\ee

\subsection{Production at Reheating}

Since it is very likely that the  short period of preheating  
does not fully extract all
of the energy density from the inflaton field, the Universe will enter a long
period of matter domination after preheating where the dominant 
contribution to the
energy density of the Universe is provided by the residual small
amplitude oscillations of the classical inflaton field and/or by the
inflaton quanta produced during the back-reaction 
processes.
This
period will end when the age of the Universe becomes of the order of
the perturbative lifetime of the inflaton field.  At this point the
Universe will go through a period of reheating with a reheat
temperature $T_{RH}$ given by the perturbative result in eq.~(\ref{thh}).

Let us suppose that the inflaton couples to $N_1$, either directly or
through exchange of other particles. In this case,
the inflaton decay process can generate a right-handed neutrino primordial
population.
The condition in eq.~(\ref{cond1}) is replaced by the weaker constraint
\be
M_1<m_\phi ,
\ee 
where $m_\phi$ is the inflaton mass.

The fate of the right-handed neutrinos produced by the inflaton decay
depends on the parameter
choice.
If $M_1<T_{RH}$ and 
$m_1\simgt 10^{-3}$~eV, the Yukawa couplings are strong enough to bring
$N_1$ into thermal equilibrium, and leptogenesis 
can proceed as in the
usual scenario described in sect.~\ref{secter}. 

Let us now assume that
the Yukawa couplings are much smaller, and that the right-handed neutrino
decay temperature in eq.~(\ref{tstar}) satisfies $T_*<T_{RH}$, {\it i.e.} $m_1<
(T_{RH}/M_1)^2 10^{-3}$~eV. After reheating,
the $N_1$ behave like 
frozen-out, non-thermal,
relativistic particles with typical energy $E_{N_1}\simeq m_\phi /2$.
The $N_1$ population will become non-relativistic at a temperature
$T_{NR}=T_{RH} M_1/E_{N_1}$. At this moment, the energy of the Universe is
shared between the radiation and the $N_1$ component, with a ratio
of the corresponding energy densities which
has remained constant between $T_{RH}$ and $T_{NR}$,
\be
\left. \frac{\rho_{N_1}}{\rho_R} \right|_{T=T_{NR}}
= \left. \frac{\rho_{N_1}}{\rho_R} \right|_{T=T_{RH}}
\ee
\bea
\left. \rho_R \right|_{T=T_{RH}}= \frac{\pi^2}{30}g_*T_{RH}^4&&~~~~
\left. \rho_{N_1} \right|_{T=T_{RH}}=E_{N_1} n_{N_1}=\frac{m_\phi}{2}B_\phi
n_\phi .
\eea
Here $n_\phi$ is the inflaton number density just before decay, obtained
by requiring energy conservation
\be
\left. n_\phi \right|_{T=T_{RH}} = \frac{\pi^2g_*T_{RH}^4}{30m_\phi
(1-B_\phi /2)}, \label{sopra}
\ee
and $B_\phi$ describes the average number of $N_1$ produced in 
a $\phi$ decay.
Below $T_{NR}$, the $N_1$ density
red-shifts like matter and eventually dominates the Universe at a temperature
\be
T_{dom} =\frac{B_\phi}{(1-B_\phi /2)}\left( \frac{M_1}{m_\phi} \right) 
T_{RH}.
\ee
Therefore, if
\be
m_1>\left( \frac{B_\phi}{1-B_\phi /2}\right)^2 
\left( \frac{T_{RH}}{10^{10}~{\rm GeV}}\right)^2
\left( \frac{10^{13}~{\rm GeV}}{m_\phi}\right)^2 1\times 10^{-9}~{\rm eV} ,
\label{disug}
\ee
then $T_*>T_{dom}$ and $N_1$ decays before dominating. In this case, the
baryon asymmetry is determined to be
\be
\frac{n_B}{s}=\epsilon a \frac{n_{N_1}}{s}=
\frac{3~\epsilon a~B_\phi ~T_{RH}}{4(1-B_\phi /2)m_\phi}.
\label{bars}
\ee
If the inequality~(\ref{disug}) is 
not satisfied, $N_1$ matter-dominates the Universe
and we recover the baryon asymmetry result in eq.~(\ref{baune}).

A necessary condition to be satisfied is that lepton-number violating
interactions mediated by $N_i$ ($i=1,2,3$) 
exchange are out of equilibrium at the
temperature of $N_1$ decay,
\be
\Gamma_{\Delta L}=\frac{4}{\pi^3}G_F^2m_i^2T^3 < H ~~~~{\rm at}~T=T_*,
\ee
where $T_*$ is given in eq.~(\ref{tstar}). This implies
\be
m_1 < \left( \frac{10^{12}~{\rm GeV}}{M_1}\right)^{2/5} 0.1~{\rm eV}~~~~{\rm
and}~~~~m_1 < \left( \frac{10^{12}~{\rm GeV}}{M_1}\right)^2
\left( \frac{2\times 10^{-3}~{\rm eV^2}}{\sum_{i=2,3} m_i^2}\right)^2 
2~{\rm eV}.
\label{lim1}
\ee

Finally, we discuss the case $T_*>T_{RH}$, {\it i.e.} $m_1>
(T_{RH}/M_1)^2 10^{-3}$~eV, in which $N_1$ decays immediately after it
is produced.
In this case, the baryon asymmetry is still given by eq.~(\ref{bars}), but
the out-of-equilibrium condition of lepton-violating interactions has to
be imposed at $T=T_{RH}$. Therefore, eq.~(\ref{lim1}) is replaced by
\be 
m_i < \left( \frac{10^{10}~{\rm GeV}}{T_{RH}}\right)^{1/2} 3~{\rm eV},~~~~
i=1,2,3.
\ee
The combination of the bounds shows that lepton-violating interactions
do not give severe constraints on the parameters.

\subsection{Production at Preheating}

As we have seen in sect.~2, right-handed neutrinos are efficiently produced in
a non-thermal state during the preheating stage. In our numerical studies we
have tacitly assumed that the superheavy right-handed neutrinos were stable. Of
course, the parametric resonance is affected by a  nonvanishing decay width of
the $N_1$. However, contrary to what happens for bosons where the presence of a
large  decay width  removes the   particles from the resonance bands  rendering
the   preheating less efficient \cite{krt},  for fermions the presence of a
decay width might be even beneficial. Indeed,  for stable right-handed
neutrinos the distribution function $n(k)$ is rapidly saturated to unity and
further particle production is Pauli-blocked. However, if the decay width is
large enough, the right-handed neutrinos may be  produced at each inflaton
oscillation when $m(t)\simeq 0$ and then   decay right away. This will  give
rise to a certain amount of lepton asymmetry at each inflaton oscillation until
the condition in eq.~(\ref{frees-out}) is met; the lepton asymmetry would be 
generated in a cumulative way. 
Strictly speaking, however, the numerical calculation of fermion preheating
presented in sect.~2
applies only to the case in which the right-handed neutrinos have a
decay lifetime larger  than the  typical time-scale of the inflaton oscillation
$m_\phi^{-1}$ 
\be 
\label{condition} 
\Gamma_{N_1}\simlt m_\phi ~~~~\Rightarrow
~~ m_1<\left( \frac{10^{15}~{\rm GeV}}{M_1} \right)^2 \left(
\frac{m_\phi}{10^{13}~{\rm GeV}}\right) 8\times 10^{-3}~{\rm eV}. 
\ee 

The right-handed neutrinos produced during preheating may
annihilate into inflaton quanta. This back-reaction will render the final
right-handed neutrino abundance smaller and therefore leptogenesis more
difficult. Imposing that the back-reaction is negligible requires $\Gamma_A\sim
n_{N_1}\sigma_A  \simlt m_\phi$, 
where $\sigma_A\sim g^4/(4\pi M_1)^2$. Since the
energy spectrum  of the  right-handed neutrinos is dominated by the maximum
momentum generated at the last inflaton oscillation,   
we assume
that the number density of
right-handed neutrinos is equal to the freeze-out value
$n_{N_1}\simeq M_1 m_\phi^2/6\pi^2$, see eq.~(\ref{frac1}),
and we obtain
\be 
\Gamma_{A}< m_\phi ~~~~\Rightarrow
~~q< 10^{13} \left( \frac{M_1}{10^{15}~{\rm GeV}}\right)^{1/2}.
\label{paf1}
\ee 

One should also be sure that the number
density of the right-handed neutrinos is not depleted by self-annihilations
before they decay. This requires 
\be
\Gamma_A< \Gamma_{N_1} ~~~~\Rightarrow
~q< 10^{12} \left( \frac{m_1}{10^{-4}~{\rm eV}}\right)^{1/2}
\left( \frac{M_1}{10^{15}~{\rm GeV}}\right)^{3/2}.
\label{paf2}
\ee

Suppose now that the right-handed neutrinos $N_1$  decay before the 
inflaton energy density is transformed into radiation by perturbative 
processes. This occurs when $\Gamma_{N_1}> \Gamma_\phi$, which implies
\be
m_1>\left( \frac{T_{RH}}{M_1} \right)^2 1\times 10^{-3}~{\rm eV}.
\label{occu}
\ee
A crucial point 
is that, after the generation of non-thermal
right-handed neutrinos at the preheating stage, the ratio of the energy
densities of $N_1$ and inflaton quanta
remains constant: $\rho_{N_1}/\rho_\phi=(\rho_{N_1}/\rho_\phi)_{ph}$, where
$(\rho_{N_1}/\rho_\phi)_{ph}$ is the ratio generated at the preheating stage. 
Since the energy density of the Universe is dominated by inflaton
oscillations, after preheating we obtain
\be
\rho_{N_1} = \left(\frac{\rho_{N_1}}{\rho_\phi}\right)_{ph} 
\frac{3H^2\mpl^2}{8\pi}.
\label{spitt}
\ee
Our numerical results discussed in sect.~2
indicate that, for very large $q$ and $M_1$, 
\begin{equation}
\left(\frac{\rho_{N_1}}{\rho_\phi}\right)_{ph} \sim 10^{-\,11} q 
\;\;\;\;\mbox{for}\;\;\;\;\;M_1 \simlt
\frac{q^{1/2}}{2}\,m_\phi.
\label{pitt}
\end{equation}
At $t_{N_1}\sim \Gamma_{N_1}^{-1}$ the right-handed neutrinos decay and  
the energy density $\rho_{N_1}$ is converted into a thermal bath with
temperature
\be
\frac{\pi^2}{30}g_* \widetilde{T}^4=\rho_{N_1},
\ee 
where $\rho_{N_1}$ is computed at $H=\Gamma_{N_1}$. 
Using eqs.~(\ref{spitt}) and (\ref{pitt}), we find
\be
\widetilde{T}=\left( \frac{q}{10^{10}}\right)^{1/4}
\left( \frac{m_1}{10^{-4}~{\rm eV}}\right)^{1/2}
\left( \frac{M_1}{10^{15}~{\rm GeV}}\right) 2\times 10^{14}~{\rm GeV}.
\ee
Before the inflaton decay,
this thermal bath never dominates the energy density of the Universe since
$\rho_{N_1}\ll \rho_\phi$ at $H\simeq 
\Gamma_{N_1}$ and the energy density in the
inflaton field $\rho_\phi$  is red-shifted away more slowly than the radiation.
Notice that at this time the asymmetry is still in the form of lepton number
since the sphalerons which are responsible for converting the lepton asymmetry
into baryon asymmetry are still out-of-equilibrium at $T=\widetilde{T}$. 
One might be worried that, since $\widetilde{T}$ is
usually larger than $T_{RH}$, too many gravitinos are produced at
the stage of thermalization of the decay products of the right-handed neutrino.
However, one can estimate the ratio $n_{3/2}/s$ after reheating to be
of the  order of $ 10^{-15}(q/10^{10})^{3/2}\left(T_{RH}/10^{10}\: {\rm
GeV}\right)$, which is quite  safe. 
However, we have to require that the lepton-number violating processes within
the thermal bath at temperature $\widetilde{T}$  are out-of-equilibrium in
order not to wash out the lepton asymmetry generated by the right-handed
neutrino decays. Therefore, we demand that 
\be
\Gamma_{\Delta L}=\frac{4}{\pi^3}G_F^2m_i^2T^3 < H ~~~~{\rm 
at}~T=\widetilde{T}.
\ee
This implies
\bea m_1 &<&\left(\frac{10^{10}}{q}\right)^{3/10} \left( 
\frac{10^{15}~{\rm GeV}}{M_1}\right)^{2/5}
1\times 10^{-2}~{\rm eV},  \\
 m_1 &<&\left(\frac{10^{10}}{q}\right)^{3/2} \left( 
\frac{10^{15}~{\rm GeV}}{M_1}\right)^2
\left( \frac{2\times 10^{-3}~{\rm eV^2}}{\sum_{i=2,3} m_i^2}\right)^2 
6\times 10^{-5}~{\rm eV}. \label{fott}
\eea

At $H\simlt \Gamma_{N_1}$, the ratio $n_L/n_\phi$ keeps constant
until the time $t_\phi\sim \Gamma_\phi^{-1}$ when the inflaton 
decays and the energy density in the inflaton field $\rho_\phi(t_\phi)$ is 
transferred to radiation. 
After reheating
we obtain the following
lepton asymmetry to entropy density ratio 
\be
\frac{n_L}{s} =\frac{3n_L\,T_{RH}}{4\rho_\phi(t_\phi)}
= \frac{3~\epsilon ~T_{RH}}{4~M_1} 
\left(\frac{\rho_{N_1}}{\rho_\phi}\right)_{ph}.
\ee
Using eq.~(\ref{pitt}), the baryon asymmetry can be expressed as
\begin{equation} \label{prelep}
\frac{n_B}{s}=  \epsilon \left(\frac{T_{RH}}{ 10^{10} \,\mbox{GeV}}\right)
\left( \frac{10^{15} \;\mbox{GeV}}{M_1} \right) 
\left(\frac{q}{10^{10}}\right)3\times 10^{-7}.
\end{equation}

The result in eq.~(\ref{prelep}) shows that the preheating production
mechanism can lead to a successful leptogenesis even for $M_1$ as large as
$10^{15}$~GeV, if $\epsilon q \sim 10^{6}$. 
Values of $q$ as large as $10^{10}$ correspond to a perturbative coupling
between $N_1$ and the inflaton $g^2\simeq 0.4$ and are compatible with
the constraints in eqs.~(\ref{paf1}) and (\ref{paf2}).
Values
of $\epsilon$ of the order of $10^{-4}$ are quite large, but can
be attained with realistic neutrino mass matrices. For values of
$M_1$ so close to the GUT scale, we expect that all the right-handed
neutrino Majorana masses are comparable in size. Moreover, the atmospheric
neutrino results, together with the requirement of perturbative Yukawa
couplings, indicate that at least one Majorana mass is less than about
$8\times 10^{15}$~GeV. This situation of comparable Majorana masses and
some large Yukawa couplings naturally leads to large values of $\epsilon$.
Also, notice that the out-of-equilibrium condition is automatically
satisfied in the preheating scenario for all 3 right-handed neutrinos,
in a large range of parameters.

Let us finally consider the case in which $N_1$ decays after the reheating
process and the inequality (\ref{occu}) is not satisfied. Again, it is
important to establish whether $N_1$ dominates the Universe before decaying.
Since the inflaton energy density is converted into radiation and the
$N_1$ are non-relativistic, the temperature at which $\rho_{N_1}$ dominates
is given by
\be
T_{dom} =\left(\frac{\rho_{N_1}}{\rho_\phi}\right)_{ph} T_{RH} .
\ee
Therefore, for $T_*>T_{dom}$, {\it i.e.} for
\be
m_1>\left(\frac{T_{RH}}{M_1}\right)^2 \left(\frac{q}{10^{10}}\right)^2 
10^{-7}~{\rm eV},
\ee
the estimate for the baryon asymmetry in eq.~(\ref{prelep})
is still valid. Otherwise, 
for $T_*<T_{dom}$,
we obtain the result in eq.~(\ref{baune}). Notice that, in this case,
the constraint $T_*<T_{RH}$ is automatically satisfied.

\subsection{Comparison of the Different Production Mechanisms}

We want to compare here the different mechanisms discussed in this section
for leptogenesis from $N_1$
decay. 
We summarize their most important features and estimate 
the size of the CP-violating parameter $\epsilon$ necessary
to reproduce the observed baryon asymmetry.

{\bf Thermal production.} This is the right-handed neutrino
production mechanism 
usually considered in the literature for
conventional leptogenesis. 
In the range $10^{-5}~{\rm eV}<m_1<10^{-2}~{\rm eV}$ 
(or $10^{-6}~{\rm eV}<m_1<10^{-2}~{\rm eV}$ in the case of the minimal
supersymmetric model)
and for $M_1<T_{RH}$,
the thermal production of unstable
$N_1$ is efficient, and values of $\epsilon$ in the range $10^{-7}$--$10^{-5}$
can account for
the present baryon asymmetry.
The boundaries of the allowed region of neutrino mass parameters
are determined as follows. For small values
of $m_1$,
the $N_1$ production rate is suppressed and larger values of $\epsilon$
are required.
For large $m_1$, the Yukawa couplings 
maintain the relevant processes in thermal equilibrium for longer times, 
partially erasing the produced asymmetry.
The values of $M_1$ are limited by the reheat temperature after inflation
$T_{RH}$, which in turn is bounded by
cosmological gravitino considerations to be below $10^8$--$10^{10}$~GeV.

{\bf Production at reheating.}
If $N_1$ is directly or indirectly coupled to the inflaton, the decay of 
the small amplitude
oscillations of the classical inflaton field at the time of reheating can
produce a right-handed neutrino population. This production mechanism
enables us 
to extend the leptogenesis-allowed region to neutrino mass parameters
which correspond to non-thermal $N_1$ populations. In particular, $M_1$ can be
as large as $m_\phi \simeq 10^{13}$~GeV.
The observed baryon asymmetry is reproduced for
$\epsilon \sim 10^{-6} 
(10^{10}~{\rm GeV}/T_{RH})(10^{-1}/B_\phi )$, where $B_\phi$ is the 
average number of $N_1$ produced by a single inflaton decay. 

{\bf Production at preheating.}
The non-perturbative decay of large inflaton oscillations during
the preheating stage can produce a non-thermal population
of very massive right-handed neutrinos. In this case, the range of $M_1$
can be extended to values close to the GUT scale, while $m_1$ is bounded
by the condition that lepton-number violating interactions are
out of equilibrium after the $N_1$ decay, see eq.~(\ref{fott}). 
A successful leptogenesis 
requires 
$\epsilon \sim 10^{-4} (10^{10}~{\rm GeV}/T_{RH})(M_1/10^{15}~{\rm 
GeV})(10^{10}/q)$, where $q$ is related to the initial inflaton configuration
and is defined in eq.~(\ref{q}). 

In conclusion, leptogenesis provides  an interesting and simple way to
explain the present cosmic baryon asymmetry. The study of the different
mechanisms in which it can be realized provides us with precious
information on the neutrino mass parameters and the early history of the
Universe.

\vskip1cm

{\bf Acknowledgments.} The work of M.P. is partially supported by the
EEC TMR network ``Beyond the Standard Model", contract no. FMRX-CT96-0090.

\end{document}